\documentclass{aa}
\usepackage{graphicx}
\usepackage{natbib}
\usepackage{amssymb}
\usepackage{amstext}
\usepackage{latexsym} 
\usepackage[sumlimits]{amsmath}
\newcommand{\Ha}{H$\alpha$}			
\newcommand{\NII}{[N{\sc ii}]}			
\newcommand{\HI}{H{\sc i}}			
\newcommand{\amin}{$^{\prime}$}		        
\newcommand{\Msolar}{\mbox{\,$M_\odot$}}        

\begin{document}
   \title{The H$\alpha$ Galaxy Survey \thanks{
Based on observations made with the Jacobus Kapteyn Telescope operated 
on the island of La Palma by the Isaac Newton Group in the Spanish 
Observatorio del Roque de los Muchachos of the Instituto de Astrof\'\i sica 
de Canarias
   }}
   \subtitle{IV. Star formation in the local Universe}
   \author{P.~A. James 
   \inst{1},
    J.~H. Knapen
    \inst{2},
     N.~S. Shane
    \inst{1},
    I.~K. Baldry
    \inst{1} and 
    R.~S. de Jong
    \inst{3}
    } \offprints{P. A. James} 
          \institute{Astrophysics Research
	  Institute, Liverpool John Moores University, Twelve Quays
	  House, Egerton Wharf, Birkenhead CH41 1LD, UK \\
	  \email{paj@astro.livjm.ac.uk} 
	  \and Instituto de Astrof\'isica de Canarias,
               E-38200, La Laguna, Spain\\
	  \and Space Telescope Science Institute, 
          3700 San Martin Drive, Baltimore, MD 21218, USA\\
          }
          \date{Received ; accepted }

\abstract
{}
{We present an analysis of the star formation properties of
field galaxies within the local volume out to a recession velocity
limit of 3000~km~s$^{-1}$.}  
{A parent sample of 863 star-forming galaxies is used to calculate a
$B$-band luminosity function.  This is then populated with star formation
information from a subsample of 327 galaxies, for which we have
H$\alpha$ imaging, firstly by calibrating a relationship between
galaxy $B$-band luminosity and star formation rate, and secondly by a
Monte Carlo simulation of a representative sample of galaxies, in
which star formation 
information is randomly sampled from the observed subset.}
{The total star formation rate density of the local Universe is found
to be between 0.016 and 0.023~$M_{\odot}$~yr$^{-1}$~Mpc$^{-3}$ with the
uncertainties dominated by the internal extinction correction
used in converting measured H$\alpha$ fluxes to star formation 
rates.  If our
internally derived $B$-band luminosity function is replaced by one
from the Sloan Digital Sky Survey blue sequence, the star formation
rate densities are $\sim$ 60$\%$  of the above values.  We also calculate
the contribution to the total star formation rate density from
galaxies of different luminosities and Hubble $T$-types.  The largest
contribution comes from bright galaxies with $M_B\sim$--20~mag, and the
total contribution from galaxies fainter than $M_B=$--15.5~mag is less
than 10\%. Almost 60\% of the star formation rate density comes from
galaxies of types Sb, Sbc or Sc; 9\% from galaxies earlier than Sb and
33\% from galaxies later than Sc.  Finally, 75 - 80\% of the total star
formation in the local Universe is shown to be occurring in disk
regions, defined as being $>$1~kpc from the centres of galaxies.}
{The star formation rate density estimates found here are consistent
with values from the recent literature using a range of different 
star formation indicators. Even though they are numerous, dwarf 
galaxies contribute little to the
star formation in the local Universe, and the bulk of the star
formation takes place in $L^*$ spirals.}


\authorrunning{James et al.}
\titlerunning{H$\alpha$ Galaxy Survey. IV.}
\maketitle
%
\section{Introduction}

In spite of many studies of the star formation (SF) process in all
types of galaxy, and of the variation of the total SF rate (SFR) as a
function of cosmological look-back time, many questions remain.  In
particular, a full census of the SF activity across all types of field
galaxies, including the faintest star-forming dwarfs, is yet to be
completed.  The \Ha\ Galaxy Survey (\Ha GS) is a study of the SF
properties of galaxies in the local Universe, using
fluxes in the \Ha\ line to determine the total rates and
distributions of SF within the selected galaxies, as a contribution to
just such a census.  The observations cover 334 galaxies, which sample
all star-forming spiral and dwarf galaxy types (S0a - Im), and the
galaxies have recession velocities less than 3000~km~s$^{-1}$.  All
galaxies were observed with the 1.0 metre Jacobus Kapteyn Telescope
(JKT), part of the Isaac Newton Group of Telescopes (ING) situated on
La Palma in the Canary Islands.  The selection and the observation of
the sample are discussed in \citet{paper1}, hereafter Paper I.

Since \Ha GS includes a large number of dwarf galaxies, which are only
detectable at very small distances, the sample as it stands is not
volume-limited.  However, the galaxy selection was performed according
to well-defined criteria in terms of galaxy apparent magnitudes,
diameters and recession velocities, and so it is possible to calculate
incompleteness corrections for galaxies of all types, from the
fraction of the total surveyed volume in which they could lie and
still satisfy the selection criteria.  By applying such corrections,
volume density statistics, such as luminosity functions, can be calculated.  
This is the primary aim of the present paper.
This analysis will then enable the calculation of the average SFR
density (in units of $M_{\odot}~$yr$^{-1}~$Mpc$^{-3}$) due to all galaxies, and
the same quantity
subdivided into the contributions from different categories of galaxy.

Thus the key scientific questions we will address are as follows:\\
How does the total SFR density derived from the \Ha GS sample compare
with other local (but generally larger-scale) estimates, and with
estimates at intermediate and high redshifts (recently reviewed by 
\citealt{hopk06})?\\
What fraction of the SFR density is contributed by galaxies as a function of
their morphological type, and as a function of their optical
luminosity?\\
What fraction of the SFR density occurs within the central bulge
regions of spiral galaxies, and what fraction in disks?\\

This approach is complementary to that adopted by other surveys of
SF in the low redshift Universe (e.g. \citealt{gallego95};
\citealt{brin04}) in that it
samples low-luminosity spiral and irregular galaxies more fully than
any previous study.  The SF indicator used (narrow-band CCD imaging in
the \Ha\ line) also allows good estimates of total SFRs without
the need to correct for missed flux which occurs for slit- or
fibre-based spectroscopy, and gives two-dimensional spatial information
which is not available from such spectroscopic approaches.

The structure of the current paper is as follows. After summarising
the survey data in Section 2, we derive, in Section~3, the local
($B$-band) luminosity function from the parent sample of galaxies used
for the survey.  This will be compared to luminosity functions from
the literature derived from a comparable field sample of galaxies.  In
Section~4, the local SFR density will be calculated, first by a direct
conversion of $B$-band luminosity to SFR, and secondly
using a Monte Carlo method to extrapolate the properties of the
observed sample of galaxies to the whole parent population from which
they were drawn.  In Section~5, the contribution to the total 
SFR density from galaxies of different absolute magnitudes
and types is investigated.  The split of SF between the
central 1~kpc in radius and disk regions is explored in Section~6.
Section~7 contains a comparison of the total SFR
density estimates derived here with literature values, and all results
are summarised in Section~8.  Finally, the appendix looks at how the 
morphological types used in the present analysis map onto the 
widely-discussed red and blue galaxy sequences.

\section{The galaxy sample selection}
\label{sec:sample}

The primary aim of \Ha GS was to determine the total SFR
of all galaxies in the local Universe, and to analyse the
contributions to this total from galaxies of different types and
luminosities. In particular, it was considered very important to give
detailed consideration to low-luminosity dwarf galaxies, which are
numerically the dominant galaxy population, but are often omitted from
studies of this type.  For example, the fainter galaxies are
inevitably numerically under-represented in any magnitude-limited
sample.  In this Section we describe the sample selection used to overcome
these problems, which enables us to infer the total SFR
of all galaxies from our observations of 327 galaxies (the remaining
seven galaxies were serendipitously observed objects and are thus omitted
from the analysis presented here).

The \Ha GS sample was selected from the Uppsala Galaxy Catalogue (UGC;
\citealt{nilson}).  We assume the UGC is complete within its selection
criteria, that is, all galaxies down to a limiting diameter of
1.0\amin\ at declinations greater than $-2.5$\degr.  
Selection biases in
the UGC are beyond the scope of this work (but see \citealt{dejong00}
who find incompleteness of both high and low surface brightness
galaxies in the similar ESO-Uppsala catalogue, \citealt{laub82}).

The \Ha GS selection criteria are described in detail in Paper I, but
can be summarised as follows: type S0a ($T=0$) or later; diameters
between 1.7 and 6.0 arcmin and recession velocities less than
3000~km\,s$^{-1}$. A $B$-band
  magnitude limit of 15.5 was adopted, corresponding to the limiting
  magnitude beyond which the UGC becomes significantly incomplete. A
  Galactic latitude limit of $|b|>$20\degr was also imposed to
  minimise the effects of Galactic obscuration.  
In addition, to prevent the skewing of the results
by the high density and special morphological characteristics of
galaxies in the Virgo cluster, galaxies in regions centred
on M~87 were excluded.  We investigate the effects of varying the size 
of the excluded region in section 3, but our adopted exclusion
for the initial sample selection
lead to a sample (which we refer to as the parent
sample) of 863 galaxies.  The observing
list was trimmed by removing the most highly-inclined spiral galaxies 
($a/b>$4.0)
since these suffer from the worst internal extinction effects and
yield less information on the spatial distribution of SF
than do face-on galaxies.  The assumption here is that galaxy
orientations are sufficiently random that no overall bias is
introduced by this cut, i.e., the remaining 743 face-on galaxies are
statistically representative of the full parent sample.  

Recession velocities are available on the NASA/IPAC Extragalactic
  Database (NED) for all UGC galaxies that satisfy the diameter,
  apparent magnitude and type requirements for our parent sample.
  There is one frustrating exception, UGC~4413, which is just large
  and bright enough for conclusion, but has no measured recession
  velocity on NED.  In any case, UGC~4413 is highly inclined and and
  would be excluded from the pool of galaxies considered sufficiently
  face-on for potential observation.  Overall, it can be considered
  that there is no incompleteness in the current sample due to missing
  recession velocities.

Within the time allocation, we observed a sub-sample of 327 galaxies
from the parent sample, which sampled the full range of galaxy
parameters in the parent sample.  This gave us detailed SF parameters
for each galaxy within this sub-sample.  The analysis presented in
this paper uses these parameters to supply estimates of the likely SF
properties, and the scatter in these parameters, of galaxies as a
function of type and absolute magnitude.  These distributions are then
used to populate a simulated volume of galaxies with SF properties,
where the relative numbers of galaxies of different luminosities are
drawn from a $B$-band luminosity function derived from our parent
sample. Whilst this process is likely to be significantly in error for
individual galaxies, it should give statistically reliable results
when averaged over many tens or hundreds of galaxies.

\begin{figure}
\centering
\rotatebox{-90}{
\includegraphics[height=8.7cm]{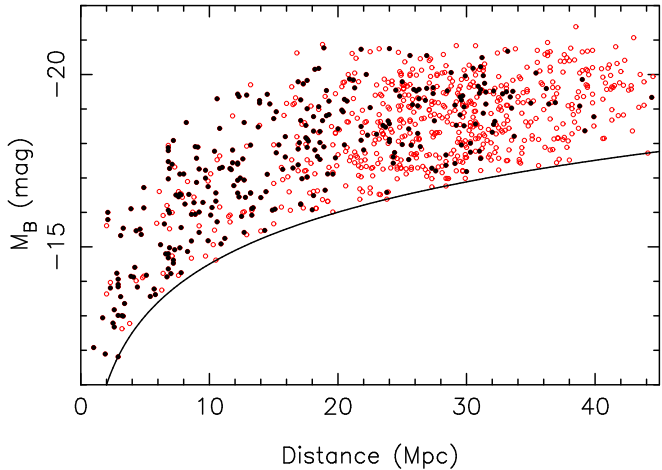}
}
\caption{The distribution of absolute blue magnitudes within the parent 
sample of 863 galaxies, as a function of distance in Mpc.  
The solid line corresponds to the apparent 
magnitude limit of $B=$15.5~mag used in selecting the galaxies. Black/filled 
symbols
denote the observed subsample, and red/open symbols the remaining unobserved 
galaxies.}
\label{fig:Mb_dist}
\end{figure}

The apparent blue magnitudes quoted by the NED for each galaxy were
used.  The NED magnitudes are generally taken from the Third Reference
Catalogue of Bright Galaxies (RC3; \citealt{rc3}).  To convert the
apparent magnitudes to absolute magnitudes, distances were calculated
for each galaxy using a Virgo-infall correction method based on that
of \citet{schechter} and an asymptotic Hubble constant of 70
km~s$^{-1}$~Mpc$^{-1}$. Such distances are very uncertain for the
closest galaxies, and, where available, distances from the literature
based on either Cepheid variables or the apparent magnitude of the tip
of the red giant branch were used in preference.
Fig. \ref{fig:Mb_dist} shows the distribution of absolute blue
magnitudes ($M_B$) within the parent sample as a function of distance
in Mpc, with the solid line showing the limiting magnitude adopted of
$B = 15.5$~mag.  The $B$ magnitudes plotted here, and used in the
following Section for the calculation of visibility volumes, are not
corrected for Galactic extinction. However, the $B$ magnitudes used
subsequently, for the calculation of the galaxy luminosity function
and the estimation of global SFRs, have corrections applied based on
the methods of Schlegel, Finkbeiner \& Davis (1998), which are also
used to correct all \Ha\ measurements (see Section 4).

\section{The $B$-band luminosity function}
\label{sec:LF}

The first stage of the analysis is to use the parent galaxy sample to
produce a $B$-band luminosity function for galaxies in the volume
sampled by this survey.  We will then use the representative subsample
observed in \Ha\ emission to populate this $B$-band function with
information on SFRs of individual galaxies, from which we can
calculate the total SFR in the local Universe.  This two-stage process
was preferred to the direct calculation of the \Ha\ luminosity
function from the 327 galaxies with \Ha\ data because the latter
sample is too small to constrain fully the form of the luminosity
function.

The luminosity function can be most directly calculated from a
volume-limited catalogue containing a fair sample of all relevant
galaxies over the range of intrinsic luminosities of interest. This is
not practicable, however, for the study of field galaxies over a wide
range of intrinsic luminosity, as is the case with the present study;
the volume over which the faintest galaxies can be seen is so small
that, far from containing a fair sample of the brightest galaxies, it
may contain none of them at all.  In this case, it is necessary to use
a statistical approach, based on the assumption that the local
Universe does not differ significantly in its galaxy content and
volume density from the Universe averaged over successively larger
volumes.  At some level, this assumption must be in error, since we
live within the density enhancement of the Local Group, and close to
the Virgo cluster and hence to the centre of the Local Supercluster
(although the Virgo cluster galaxies themselves have been removed from
the sample).  This is likely to bias the calculated volume density to
high values, but this is unavoidable and should be borne in mind when
interpreting the results.

We use a statistical method  based on the classical
procedures outlined by \cite{schmidt68} and by \cite{felton76}, to derive 
a $B$-band
luminosity function from the parent sample in our present study.
Mathematically, this is done by numerically weighting each of the
parent sample galaxies by $(1/V_{\rm m})$, where $V_{\rm m}$ is the size in
Mpc$^3$ of the volume in which the galaxy could lie and still be
included in our sample, satisfying the apparent magnitude, diameter
and recession velocity criteria. Then the luminosity function
$\phi(M_B)$ is given by

\begin{equation}
\phi(M_B) = \frac{4\pi}{\Omega}\sum_i\frac{1}{V_{\rm m}^i}.  
\label{eqn:lf}
\end{equation}

$\phi(M_B)$ is the space density of galaxies with blue magnitude
$M_B$, $V_{\rm m}^i$ is the visibility volume within which the $i$th galaxy
could lie and still be included in the sample, and the quantity $\Omega/4\pi$
is the fraction of sky covered by the sample in question.  In this
case, $\Omega$ = 2$\pi\times$0.695, once the Virgo Cluster,
Galactic plane and southern hemisphere regions have been removed.  

Each selection criterion will define a maximum volume in which a
galaxy of given properties can be observed.  The visibility volume for 
each galaxy is thus the volume which satisfies all these criteria
\citep{disney83,davies94}.  There are three selection limits
applicable to the parent sample (apparent magnitude, diameter and 
recession velocity), each of which leads to constraints
on the distance range within which a galaxy must lie to be included in the 
sample.  We adopt a limiting apparent magnitude,
$m_{\rm lim}$, of 15.5~mag, and have excluded the small number of UGC galaxies
fainter than this in defining the parent sample.  A galaxy with
absolute magnitude $M$ must therefore lie within a distance in Mpc given by
\begin{equation}
d_{\rm max,mag}\ (\text{Mpc}) = 10^{0.2(m_{\rm lim}-M-25)}.
\end{equation}
This equation applies to magnitudes uncorrected for Galactic and
internal extinction. There is no bright apparent magnitude limit so
$d_{\rm min,mag} = 0$~Mpc for all galaxies.   The lack of bright
galaxies at distances up to 16~Mpc, clearly apparent in
Fig. \ref{fig:Mb_dist}, is due to the upper diameter limit on galaxy
selection, explained in the following paragraph.

The \Ha GS sample was selected to contain galaxies with apparent
angular diameters between 1.7\amin\ and 6.0\amin.  A galaxy with an
apparent angular diameter $D$ will be excluded from the sample if its
distance, $d$, is low enough for the apparent diameter to be
greater than 6.0\amin, or if it is so far away that the apparent
diameter falls below 1.7\amin.  Thus the distance limits within which it can be
observed are

\begin{equation}
d_{{\rm min},D}\ (\text{Mpc}) = d\left(\frac{D}{6.0}\right)
\end{equation}

and

\begin{equation}
d_{{\rm max},D}\ (\text{Mpc}) = d\bigg{(}\frac{D}{1.7}\bigg{)}
\end{equation}

Galaxies which are intrinsically small and faint will only be detected
at the smallest distances due to the apparent magnitude and the
minimum apparent diameter limits.  The galaxies with the largest
intrinsic sizes can only be observed at larger distances, due to the
maximum angular diameter limit.  

The visibility volume $V_{\rm m}$ is then defined by

\begin{equation}
V_{\rm m}\ (\text{Mpc}^3) = \frac{4\pi}{3} (d_{\rm max,lim}^3-d_{\rm min,lim}^3),
\end{equation}

where $d_{\rm max,lim}$ is the smallest of $d_{\rm max,mag}$,
$d_{\rm max,D}$ and the distance corresponding to 3000~km~s$^{-1}$
in the direction of the galaxy. $d_{\rm min,lim}$ is equal to $d_{\rm min,D}$.

\begin{figure}
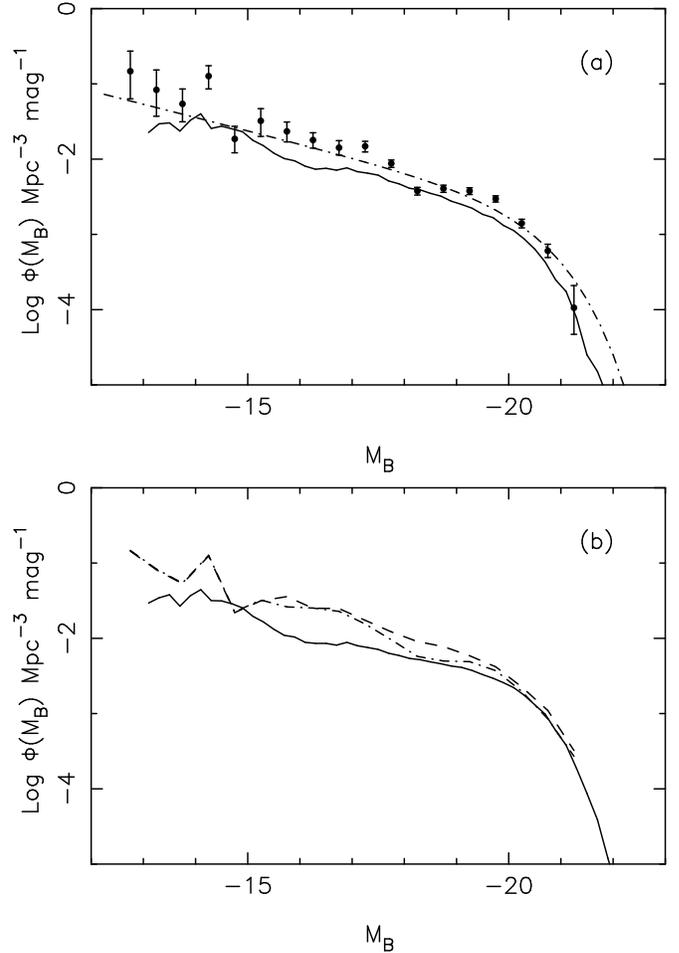

\centering
\rotatebox{-90}{
\includegraphics[height=8.7cm]{8560fg2a.ps}
\includegraphics[height=8.7cm]{8560fg2b.ps}
}
\caption{(a) The luminosity function for the entire parent sample, types
S0a -- Im, plotted as points with error bars;
the solid line shows the luminosity function from the SDSS blue 
sequence. (b) Luminosity functions including all galaxy types:  SDSS 
(solid line); this 
study with all Virgo galaxies included (dashed line); and this study with
galaxies excluded out to 10\degr from the centre of the Virgo cluster 
(dot-dashed line).}
\label{fig:parent_lf}
\end{figure}

The $B$-band luminosity function is then calculated by performing the
sum in Equation \ref{eqn:lf} over all galaxies in the parent sample.
The resulting function is shown in Fig. \ref{fig:parent_lf}.  The
error bars contain two contributions: 1~$\sigma$ Poisson errors taken
from the tables of \cite{gehr86}, which are added in quadrature to
distance errors, calculated using the method of \cite{dejong00}.  Here
we adopt distance errors of $\pm$50\% for all galaxies within 5~Mpc;
$\pm$30\% for all galaxies between 5 and 10~Mpc; $\pm$20\% for all
galaxies between 10 and 15.6~Mpc; and $\pm$15\% for all galaxies
beyond 15.6~Mpc.

Luminosity functions of field galaxies are often found to be described
well by a Schechter function \citep{schech76} (see, e.g., Efstathiou,
Ellis \& Peterson 1988):

\begin{equation}
\phi(L)dL = \phi^*(L/L^*)^{\alpha}\exp(-L/L^*)d(L/L^*),
\label{eqn:Lscht}
\end{equation}

This function 
decreases as a power law with increasing galaxy
luminosity at faint magnitudes, but cuts off sharply for galaxies
brighter than some characteristic magnitude, $M^*$, corresponding to 
$L^*$.  $\alpha$ gives
the slope of the luminosity function at the faint end and $\phi^*$
sets the overall normalisation of galaxy density.  

The luminosity function derived from our parent sample is shown
  as the points with error bars in Fig. \ref{fig:parent_lf}(a).  A
  Schechter fit was made to this luminosity function, and is shown in
  Fig. \ref{fig:parent_lf}(a) as the dot-dashed line. The values of
  $M^*$ and $\phi^*$ in this fit are poorly constrained, due to the
  small number of galaxies brighter than the turnover at $M^*$ and the
  degeneracy between these parameters.  Formally, the best-fit values
  are $M^* = -20.49 \pm 0.19$, $\phi^* = 0.00280 \pm 0.00062$, and the
  faint-end slope $\alpha = -1.44 \pm 0.06$. It should be noted,
  however, that a single Schechter function is not a good fit, due to
  the plateau in the observed function between $M_B=-$18.0 and
  --19.5~mag.

The solid line in Fig. \ref{fig:parent_lf}(a) is an estimate of the
$B$-band luminosity function for blue-sequence galaxies.  This is
derived from the Sloan Digital Sky Survey (SDSS), using the New York
University Value-Added Galaxy Catalog (NYU-VAGC;
\citealt{blanton05A}), in particular, the low-$z$ galaxy sample
\citep{blanton05B}. This sample includes galaxies with $r<17.8$~mag in
the redshift range $1000<cz<15\,000{\rm\,km\,s}^{-1}$ (14 - 212\,Mpc
assuming $H_0=70$~km~s$^{-1}$~Mpc$^{-1}$). Galaxies have been
eyeballed to check photometry and redshifts have been corrected to
distances using a local flow model for the nearest galaxies.
K-corrections are applied to bring the photometry to $z=0$.  The
NYU-VAGC absolute magnitudes were converted to $B$ magnitudes using
\begin{equation}
B = g + 0.14 + 0.37 (g-r) 
\end{equation}
which was obtained by matching Millennium Galaxy Catalogue (MGC)
galaxies \citep{liske03} to SDSS galaxies and fitting a straight line
to $B-g$ versus $g-r$ (using $z<0.1$ galaxies).  The SDSS luminosity
function was then determined by summing, in absolute magnitude bins,
\begin{equation}
  \frac{1}{V_{\rm max} C_{\rm sb}}
\end{equation}
where $V_{\rm max}$ is obtained from the NYU-VAGC and $C_{\rm sb}$ is
the surface-brightness completeness for each galaxy estimated using
table~1 of \citet{blanton05B}.  The luminosity function is then
effectively complete to a half-light surface brightness of
$\sim24{\rm\,mag\,arcsec}^{-2}$ in $r$. Counts and magnitudes were all
converted to $H_0=70$~km~s$^{-1}$~Mpc$^{-1}$.

To determine the SDSS luminosity function for late type galaxies three
cuts were considered: (i) a colour cut selecting blue-sequence
galaxies as per \citet{baldry06}; (ii) Sersic index less than 2.5 from
the NYU-VAGC; (iii) concentration index, $R90/R50$, less than 2.5 from
the SDSS pipeline.  These three definitions of the blue sequence yield
largely consistent luminosity functions, and we adopted the first,
hereafter referred to as the SDSS blue-sequence luminosity function,
to compare with that from the H$\alpha$GS parent sample.

Overall there is a fair agreement in the shape and normalisation of
the SDSS blue-sequence and H$\alpha$GS luminosity functions, but there
are two areas of significant disagreement.  The first is that the
H$\alpha$GS luminosity function is somewhat higher around $M_B=-19.0$
to $-20.5$~mag, which is significant as these galaxies dominate the
stellar mass and thus, plausibly, the SFR density of the overall
galaxy population.  Secondly, the H$\alpha$GS luminosity function
shows a somewhat steeper faint-end rise, i.e., a larger dwarf-to-giant
ratio than the SDSS function.  The possible causes of these differences
will now be investigated.

One possibility is that the higher amplitude of our luminosity
function is due to the presence of the Virgo cluster, which may still
have a significant influence even though the core region is excluded.
This possibility was explored by producing luminosity functions for
local galaxies selected from the UGC, with differently-sized regions
of the Virgo cluster removed.  For ease of comparison with literature
studies, this comparison was done for all Hubble types, i.e. with
types earlier than S0a reinstated, but otherwise the selection
criteria are as described above.  Two luminosity functions derived in
this way are shown in Fig. \ref{fig:parent_lf}(b), where the dashed
line includes all Virgo cluster galaxies, and the dot-dashed line
shows the effect of excluding all galaxies within a 10\degr\ radius
centred on M87. (5\degr, 15\degr\ and 20\degr\ radius regions were
also investigated, with similar results).  For comparison, the full
SDSS luminosity function, including all galaxy types, is shown as the
solid line.  The main effect of Virgo exclusion on the UGC luminosity
function is to depress it between $M_B$ values of --18 and --20, thus
causing the plateau mentioned above.  The bright end of the function
is then in fair agreement with the SDSS-derived luminosity function.
At $M_B$ = --20, the UGC-derived function has an a value $\phi$(--20)
of 0.0031~Mpc$^{-3}$mag$^{-1}$ with Virgo included, falling to 0.0029,
0.0027, 0.0027 and 0.0025~Mpc$^{-3}$mag$^{-1}$ for excluded regions of
radius 5\degr, 10\degr, 15\degr\ and 20\degr\ respectively.  For the
full SDSS-derived luminosity function, $\phi$(--20) is
0.0024~Mpc$^{-3}$mag$^{-1}$.  This improved agreement shows that at least part
of the bright-end discrepancy in Fig. \ref{fig:parent_lf}(a) is due to
the SDSS blue sequence definition excluding a larger fraction of
galaxies than our S0a - Im type constraint.  This is supported  by 
the analysis presented in the appendix, which shows that the blue 
sequence most nearly corresponds to types Sb or later.

The literature sample most comparable to our parent sample is that of
\citet{sant96}, who studied the luminosity and diameter functions of
galaxies within 8000~km~s$^{-1}$ selected from both the UGC and its
southern counterpart, the ESO-Uppsala Survey (\citealt{laub82}).  They
do not present Schechter function fit parameters for their luminosity
functions, citing significant biases in the sample selections which
are hard to quantify, but the $\phi$(--20) values for their
luminosity functions from both catalogues are somewhat lower than
those we derive, at 0.0019--0.0021~Mpc$^{-3}$mag$^{-1}$, after 
correction from their assumed Hubble constant of 
100~km~s$^{-1}$~Mpc$^{-1}$.

It is clear from this analysis that cosmic variance may have a
significant effect on the H$\alpha$GS luminosity function over the
volume we survey, and this will be more important at the faint
end, due to the even smaller region sampled by the faintest galaxies.
In order to quantify this faint-end difference, estimates were made
of the amplitude of the $B$-band LF at $M_B=$ --16.0, for comparison 
with literature studies.  With all galaxy
types and the Virgo cluster included, our UGC-based sample gave a value
for $\phi$(--16) of 
0.030~Mpc$^{-3}$mag$^{-1}$
falling to 
0.025~Mpc$^{-3}$mag$^{-1}$
with the exclusion of the Virgo cluster, and to
0.020~Mpc$^{-3}$mag$^{-1}$
with the additional exclusion of galaxies earlier than type S0a.
The SDSS and \citet{sant96} luminosity functions are significantly lower
in this region, with $\phi$(--16) values between  0.006 and
0.011~Mpc$^{-3}$mag$^{-1}$.  Thus cosmic variance seems likely to 
have a substantial effect at these faint magnitudes.

However, it is also true that the larger-scale surveys have
substantial incompleteness below $M_B\sim-$16~mag, and the
corresponding uncertainties make it impossible to rule out the higher
faint end counts that we find in the current sample.  Indeed, some
other studies of the local galaxy luminosity function support a faint
end slope as steep as, or steeper than, that found here for the
H$\alpha$GS parent sample.  For example, \citet{mari99}, studying an
all-sky sample of 6400 optically selected galaxies within
5500~km~s$^{-1}$, found a very steeply-rising faint end slope ($\alpha
<$ --2 for the faintest galaxies, vs $\sim$--1.1 over most of the
luminosity range studied) with the steepening at $M_B \sim-15$~mag
being ascribed to Magellanic irregular galaxies.

In light of these uncertainties in the form of the local field galaxy
luminosity function, for the remainder of this paper we will adopt the
H$\alpha$GS and SDSS blue sequence luminosity functions as limiting cases,
and use both when characterising the SF properties of the local
Universe.

\section{The local SFR density}
\label{sec:sfr-LF}

\subsection{Methods for estimating the SFR density from the $B$-band luminosity function}

The $B$-band luminosity function calculated in the previous Section
from our parent sample gives the total number of galaxies of different
luminosities per unit volume of the Universe. The next stage of this
analysis is to populate the individual galaxies making up this luminosity
function with SF information, thus enabling calculation of the total
SFR density, and the breakdown of this total by galaxy type.  The SF
information is provided by the observed subsample of \Ha GS galaxies.

Two methods are presented here.  The first uses galaxy $B$-band
luminosity as a direct proxy for SFR, with calibration factors linking
the two quantities being derived from the observed subsample, and applied
to the full parent sample.  The scatter in this conversion process is
significant, however, and there are possible concerns that taking mean
properties of galaxies may underestimate the effects of outliers in the
distributions of SF properties.  These problems are addressed with the
second method, which uses a Monte Carlo method to build multiple
models of the Universe, with galaxy types and luminosities drawn from
the parent sample, and SF information randomly sampled from galaxies
with similar properties in the observed subsample.

The \Ha\ fluxes are converted to SFR values using the calibration
  of \citet{kenn98}, which assumes a \citet{salp55} stellar initial
  mass function.

\subsection{The SFR:$L_B$ correlation}

The first of these two methods relies on the fact that $B$-band
luminosity ($L_B$) in star-forming galaxies largely comes from the
young stellar population, and hence we can expect galaxy $L_B$ values
to correlate strongly with SFRs.  This leads to a simple way to
calculate the total SFR density: we calibrate the SFR:$L_B$ relation
from the observed subsample, then use it to replace the $L_B$ of each
galaxy in the parent sample with a SFR.  Finally, the same 1/V$_{\rm max}$
analysis as was used to calculate the $B$-band luminosity function
will now result in the total SFR density.

\begin{figure}
\centering
\rotatebox{-90}{
\includegraphics[height=8.7cm]{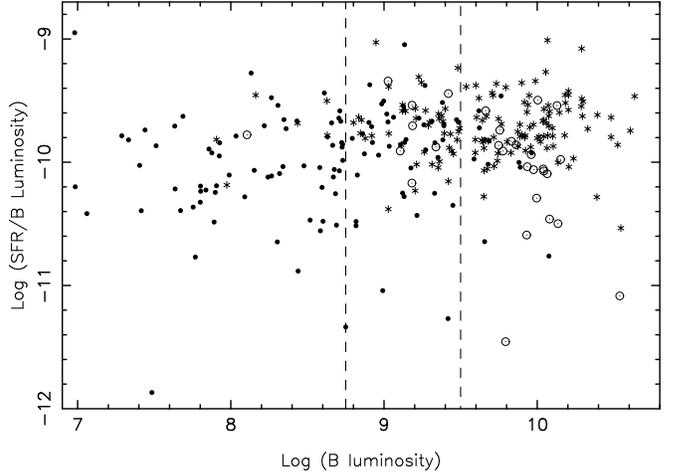}
}
\caption{The log of the SFR to $B-$band luminosity ratio,
plotted against the log of the $B-$band luminosity, for the observed galaxies.
Circled points denote early types, $T=$0 - 2; stars are intermediate
types, $T=$3 - 7; and filled circles are late types, $T=$8 - 10. }
\label{fig:sfr_v_LB}
\end{figure}

Figure \ref{fig:sfr_v_LB} shows the log of galaxy SFRs
per unit $B$-band luminosity (in solar units), plotted against the log
of the $B$-band luminosities, with different symbols showing early
Hubble types ($T$-type 0 - 2; circled points), intermediate types
($T$-type 3 - 7; stars) and late types ($T$-type 8 - 10; filled
circles).  All measured fluxes are corrected for Galactic extinction,
and the SFRs have been corrected for extinction
internal to the galaxy concerned using the type-dependent corrections
derived in Paper~II (\citealt{paper2}).  Overall,
Fig. \ref{fig:sfr_v_LB} shows no strong trend, confirming that
$B$-band luminosity can be used as a proxy for SFR,
but it is clear that, for example, the bright, early type galaxies
have a lower mean ratio than is typical of other galaxies.  Thus,
calibration factors were calculated for each of these three type bins
($T=$0 - 2, 3 - 7 and 8 - 10) and three $L_B$ bins (Faint, log($L_B$)$<$8.5;
Intermediate, 8.5$<$log($L_B$)$<$9.5; and Bright, log($L_B$)$>$9.5).
In each case, the values were calculated by dividing the sum of the
galaxy SFRs by the sum of their $B$-band luminosities in solar units,
and finally taking the log.  The resulting values are given in Table
\ref{tbl:sfr_lb}.

\begin{table*}
\begin{center} 
\begin{tabular}{c|ccc}
\hline
\hline
       & Faint & Intermediate & Bright     \cr
\hline
$T=0 - 2$   & --9.78 & --9.63 & --10.00 \cr
$T=3 - 7$   & --9.64 & --9.67 & --9.63  \cr
$T=8 - 10$  & --9.88 & --9.74 & --9.91  \cr
\hline
\end{tabular}
\caption[]{log(SFR$_{\rm TOT}$/$L_{B,{\rm TOT}}$) 
values for observed sample galaxies,
binned by luminosity and Hubble $T$-type.}
\label{tbl:sfr_lb}
\end{center}
\end{table*}

It is now simple to convert the $L_B$ values for each of the parent
sample galaxies to SFR, by multiplying by the calibration factor for
the appropriate type and luminosity bin.  Applying a $1/V_{\rm m}$
weighting and summing over the whole sample, we find the SFR density
to be 0.0237$\pm$0.0030~$\Msolar~$yr$^{-1}~$Mpc$^{-3}$.  Here the
quoted error is the random error resulting from uncertainties in the
individual galaxy \Ha\ fluxes \citep{paper1} and distances, and
Poisson errors on the numbers of galaxies used in calculating mean
binned properties.  Much larger systematic uncertainties result from
the extinction corrections applied in the conversion of \Ha\ fluxes to
SF rates, and from the luminosity function used.  These factors will
be addressed in the remainder of this section.

We will next find the SFR density that results from
using the SDSS blue sequence luminosity function.  Again the method
employed was to replace $L_B$ with SFR values from our
observed galaxy sample. However, in this case we only have the
$B$-band luminosity function, and no information on the fraction of
galaxies of each Hubble type within each $L_B$ bin. Therefore, we
calculate the (SFR/$L_B$) calibration factors simply as a function of
$L_B$ within our observed subsample.  Since the sample is not
subdivided by types, we can bin more finely in $L_B$, and so the mean
calibration factors are calculated within nine bins of width 0.4 in log
$L_B$. The SFR density from the SDSS $B$-band luminosity function
using this `direct' calculation method
0.0138$\pm0.0015$~$\Msolar~$yr$^{-1}~$Mpc$^{-3}$.

This is significantly lower than the value from the internally-derived
luminosity function, due to the overall lower level of the SDSS 
luminosity function.

\subsection{A Monte Carlo method for deriving the SFR from the $B$-band luminosity function}

There is a concern that the above method implicitly assumes a perfect
correlation between $B$-band luminosity and current SFR; this is
clearly not a valid assumption, and even within our type-luminosity bins, the
scatter in (SFR/$L_B$) is large, at 0.3~dex.  We have investigated the
possible effects of this scatter using a Monte Carlo method, producing
multiple realisations of the local Universe.  These models were
populated using the 1/$V_{\rm m}$, $T$-type, and $M_B$ information for each
of the 863 galaxies in the parent sample.  The 1/$V_{\rm m}$ value gives
the number of times each galaxy is reproduced in the model universe.
For each of these reproduced galaxies, a SFR is selected by randomly
sampling a galaxy from the observed subset with closely matching $M_B$
and identical $T$-type (we use three $M_B$ bins for each $T$-type; the
selected galaxy can be any of the observed galaxies in same bin) and
using the corresponding SFR.

\begin{figure}
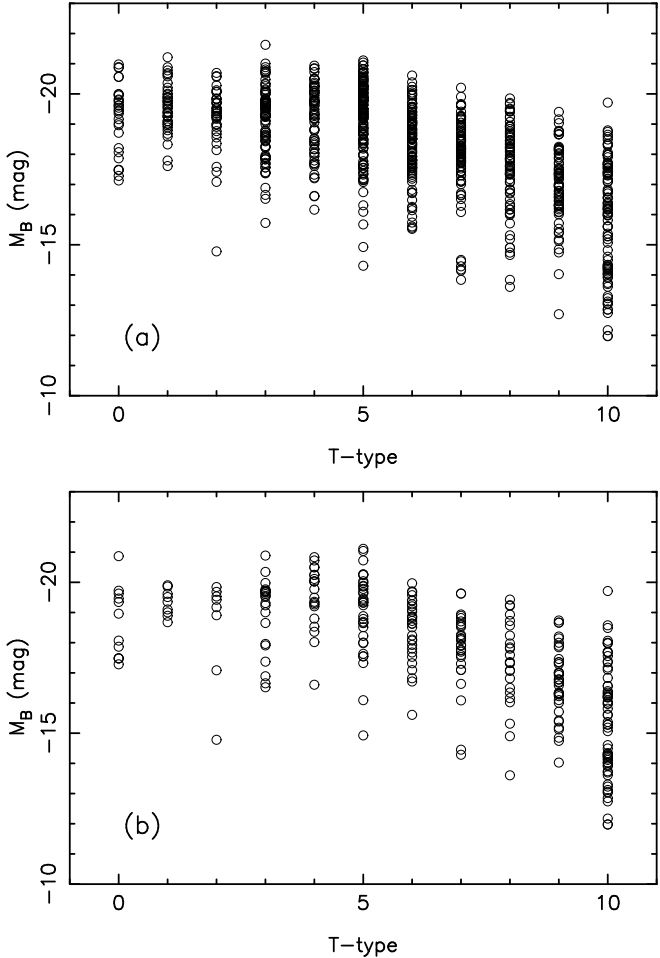

\centering
\rotatebox{-90}
{
\includegraphics[height=8.7cm]{8560fg4a.ps}
}
\rotatebox{-90}
{
\includegraphics[height=8.7cm]{8560fg4b.ps}
}
\caption{The distribution of absolute magnitudes amongst each galaxy $T$-type 
for both the parent (a) and the observed (b) samples.}
\label{fig:M_v_T}
\end{figure}

This method requires the parent and observed samples to have
well-matched properties at each $T$-type, so that every parent galaxy
can have SF information supplied from an observed galaxy with similar
characteristics.  Figure \ref{fig:M_v_T} shows the distribution of
$B$-band absolute magnitudes for each $T$-type for both parent and
observed samples, showing that indeed the distributions are
well-matched, as would be expected given that 40\% of the full sample
were observed with a good spread over all $T$-types.
Kolmogorov-Smirnov (K-S) tests applied to each galaxy type show that
the observed sample is consistent with being drawn randomly from the
parent sample.

The following results for
the total SFR density were obtained from the Monte Carlo analysis.  
Using the $B$-band luminosity function
derived from the parent sample and type-dependent extinction
corrections from Paper~II in converting \Ha\ fluxes to SFRs, the SFR 
density is 0.0229$\pm$0.0016~$\Msolar~$yr$^{-1}~$Mpc$^{-3}$.  Using the SDSS
blue-sequence $B$-band luminosity function, with the same \Ha\
extinction corrections, this value is reduced to
0.0130$\pm$0.0009~$\Msolar~$yr$^{-1}~$Mpc$^{-3}$.  In both cases, the
scatter in values found using repeated realisations of the Monte Carlo
method are less than the formal errors, which are dominated by Poisson
statistics on the size of the observed samples.  It is encouraging to
note that in both cases, the Monte Carlo method gives SFR density
values within a few percent of those found using the SFR:$L_B$
correlation.

\begin{figure}
\centering
\rotatebox{-90}
{
\includegraphics[height=8.7cm]{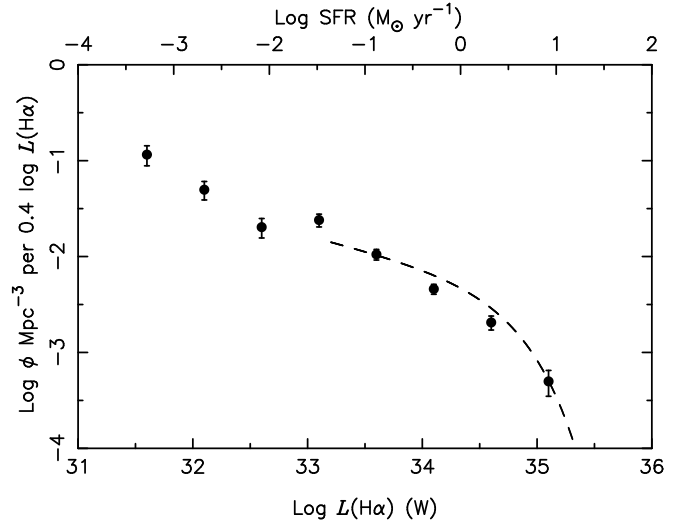}
}
\caption{
The \Ha\ luminosity function derived from our parent sample, with 
type-dependent extinction corrections. The dashed line shows \Ha\ luminosity
function of \citet{gallego95}. The horizontal scale is given both in terms 
of Log(\Ha\ )luminosity, in Watts (bottom scale), and Log(SFR), in solar 
masses per year (top scale).
}
\label{fig:half}
\end{figure}

The \Ha\ luminosity function derived from our $B$-band luminosity
function, and with the type-dependent extinction corrections, is shown
in Fig. \ref{fig:half}.  For comparison, the \Ha\ luminosity function
of \citet{gallego95} is overplotted as a dashed line, showing good
agreement in overall level over the range of SFR probed by
\citet{gallego95}.

We next used the Monte Carlo technique to evaluate the effect of
different \Ha\ extinction corrections on the derived SFR density,
using the $B$-band luminosity function from the parent sample of the
present study.  Replacing the type-dependent correction with the
constant 1.1~mag correction proposed in \cite{kenn98} gives a SFR
density of 0.0203$\pm$0.0014~$\Msolar~$yr$^{-1}~$Mpc$^{-3}$. A third
option is the $M_R$-dependent extinction correction derived by
\cite{helm04} based on the study of a 21~cm-selected galaxy sample (we
also use their \NII\ contamination correction, for consistency).  This
gives rather lower SFRs for a given \Ha\ flux, and a
total SFR density of 0.0159$\pm$0.0012~$\Msolar~$yr$^{-1}~$Mpc$^{-3}$.
Thus the systematic uncertainties associated with internal extinction
corrections are $\pm20\%$, and much larger than the formal statistical
errors.


\section{The contribution of galaxies of different types and luminosities
to the local SFR density}

\subsection{SFR density as a function of galaxy luminosity}

Using the Monte Carlo method described in the previous Section, it is
simple to evaluate the contributions to the local SFR
density from galaxies of different absolute magnitudes.  This is of
interest given the relatively large number of dwarf galaxies in the 
present sample, which could potentially offset the low individual
SFRs of these dwarfs.

\begin{figure}
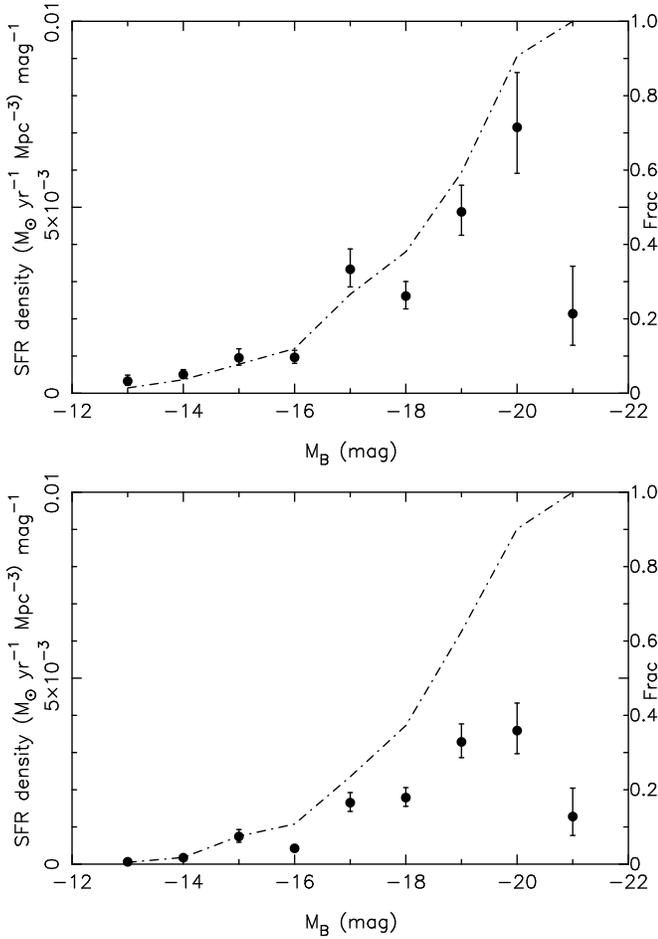

\centering
\rotatebox{-90}
{
\includegraphics[height=8.7cm]{8560fg6a.ps}
\includegraphics[height=8.7cm]{8560fg6b.ps}
}
\caption{SFR density as a function of absolute blue magnitude found
using the points in the $B$-band luminosity function shown in
Fig. \ref{fig:parent_lf} (top); and using
the SDSS $B$-band luminosity function (bottom). The cumulative 
fraction of the SFR
density is shown by the dot-dashed line in each plot; the scale for this 
fraction is marked along the right-hand axis.}
\label{fig:sfrfunc}
\end{figure}

This calculation was done using both the $B$-band luminosity function
derived from our parent sample, and that of the SDSS blue sequence.
The results are shown in Fig. \ref{fig:sfrfunc}, from which it is
clear that the local SFR density is dominated by
bright spiral galaxies with $M_B \sim -19$ - $-20$~mag, regardless of which
luminosity function is used.  Even with our relatively dwarf-rich
luminosity function, only 8\% of the SF in the local
Universe is occurring in dwarf galaxies fainter than $M_B=$--15.5~mag;
using the SDSS blue-sequence luminosity function this falls to 6\%.

\subsection{SFR density as a function of galaxy type}

We can also use our Monte Carlo method to explore the fraction of the
total SFR density contributed by galaxies of different
types.  This uses a similar method to that in the previous Section
(5.1), but makes use of the $T$-type information that is available for all
galaxies in the parent sample used in this simulation.
 
\begin{figure}
\centering
\rotatebox{-90}{
\includegraphics[height=8.7cm]{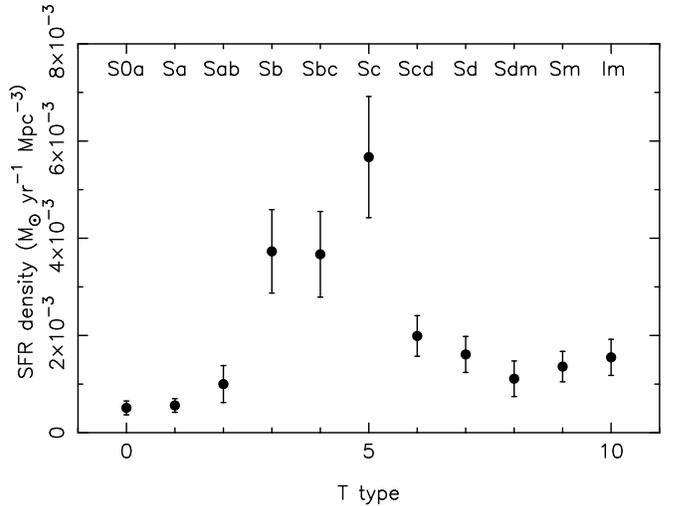}
}
\caption{SFR density as a function of galaxy morphological type, with 
galaxy numbers drawn from our $B$-band luminosity function. 
}
\label{fig:tsfrfunc}
\end{figure}

The results are shown in Fig. \ref{fig:tsfrfunc}, which  shows that
the galaxies contributing the most to the local SF are those
classified as Sb - Sc.  Early type spirals provide only a small
contribution in the present epoch, in spite of including many of the
highest mass spiral galaxies. It is interesting to note the large step
in SFR density between $T=$2 and 3 (Sab and Sb), since these types
also straddle the transition between the SDSS red and blue sequences,
as explained in Appendix A. Late-type Magellanic galaxies of types Sm
\& Im ($T=$ 9 and 10), which are the most numerous galaxies in the
field, contribute $\sim$13\% (our $B$-band luminosity function) or
$\sim$10\% (SDSS luminosity function) of the local SFR density.

\section{Disk vs central SF}

A final study, made possible by the spatial coverage and resolution
of the \Ha\ imaging technique, is to separate out SF taking
place in central and disk regions of the galaxies observed.  In this 
first analysis, we do not attempt a bulge-disk deconvolution of the
galaxies, but simply define the central region (`bulge') as being the
central 1~kpc in radius (following, e.g., \citealt{gala06}), with `disk'
SF being traced by the \Ha\ emission lying outside this radius.
For inclined galaxies, elliptical apertures with semi-major axis 1~kpc
and axial ratios matching those of the outer disk were used, as discussed
in Paper~I.  Disk and bulge fractions of SF were 
calculated in this way for all galaxies of types S0a - Sdm in our observed 
sample; the latest-type galaxies (Sdm, Sm and Im) are deemed to be bulge-free,
and hence the disk fraction for these galaxies is 100\% by definition.

One reason for doing this analysis is that the disk emission thus
defined can confidently be interpreted as being dominated by SF.  
The bulge component can potentially be contaminated by
AGN-related emission, and this process puts an upper limit on the
potential over-estimate of our SFR density values from
this cause.  Similarly, central regions can be affected by high
\NII/\Ha\ ratios (Paper~II and references therein), which can
again lead to over-estimates of SFRs. On the other
hand, by interpreting all the emission (bulge and disk) as being due
to \Ha\ from SF, we can derive an estimate of the relative
rates of growth of the stellar mass of the two components in the
current epoch, bearing in mind that the bulge estimates are likely to 
be upper limits.

We find the following disk percentages of the local SFR density as a
function of galaxy $T$-type: $T$=0, 63\%; $T$=1, 54\%; $T$=2, 73\%;
$T$=3, 64\%; $T$=4, 76\%; $T$=5, 83\%; $T$=6, 81\%; $T$=7, 78\%.
Types $T$=8, 9 and 10 are defined to have disk SF fractions of 100\%
(pure disk systems).  The total fraction of the SFR density occurring
within disks is 75\% for types 0-7; this increases to 80\% when
including the SF of the later type galaxies in the disk total.  This
demonstrates that the major part of the SF activity is occurring in
disk regions, and hence that our estimates of the total SFR density
cannot be significantly affected by an AGN contribution.  The bulge
fraction of line emission for the earliest-type galaxies {\em is}
high, however, at almost 50\% of the total.  This may imply that these
galaxies are still building the stellar mass in their central regions
at a significant rate, or it may be the result of substantial
contamination from AGN or other non-SF related emission.

A fuller investigation of the spatial distribution of line emission
within these galaxies will be described in a later paper in this series
(James et al., in preparation).

\section{Comparison with other determinations of SFR density}

We finally compare our estimates of the local SFR
density with estimates of this same parameter from the literature,
where the latter studies use a variety of field galaxy samples and
tracers of galaxy SFRs.  The present study found
values of this parameter ranging from 0.0159 to
0.0229~$M_{\odot}$~yr$^{-1}$~Mpc$^{-3}$ using the internally-derived
$B$-band luminosity function, with the range reflecting the different
internal extinction corrections used.  Using the $B$-band luminosity
function for the SDSS blue sequence gave a lower estimate, of
0.0130~$M_{\odot}$~yr$^{-1}$~Mpc$^{-3}$.

Literature results in this area have recently been surveyed by
\citet{hani06}, who quote results from several surveys, with and
without extinction corrections, and with and without a redshift
correction, since the surveys quoted have mean effective redshifts
from 0.01 to 0.20.  The numbers presented here include both extinction
correction and the redshift correction to bring the SFR densities to
an equivalent at $z=0$.

Two results come from the Universidad Complutense de Madrid survey of
optically-selected emission line galaxies, with \citet{gallego95}
finding a value of 0.0112$\pm0.001$ $M_{\odot}$ yr$^{-1}$ Mpc$^{-3}$,
and \citet{pere03} 0.0229$\pm0.001$ $M_{\odot}$ yr$^{-1}$ Mpc$^{-3}$.
The $I$-band selected Canada-France Redshift survey of
\citet{tresse98} resulted in a SFR density value of
0.0186$\pm0.0032$~$M_{\odot}$ yr$^{-1}$~Mpc$^{-3}$, corrected from an
effective mean redshift of 0.2 to $z=0$.  \citet{sull00} use a
UV-selected sample of galaxies over quite a wide redshift range (0 -
0.4) to derive a SFR density of 0.01$\pm0.001$ $M_{\odot}$ yr$^{-1}$
Mpc$^{-3}$, although this low value is completely dependent on the
redshift correction adopted by \citet{hani06}.  Easily the most
extensive study in terms of numbers of galaxies included, is the
SDSS-based study of \citet{brin04}, who find a SFR density of
0.0219$^{+0.0016}_{-0.0048}$~ $M_{\odot}$ yr$^{-1}$ Mpc$^{-3}$.
Finally, for their own, \HI - selected field sample, \citet{hani06}
find a SFR density of 0.0155$\pm0.001$~$M_{\odot}$ yr$^{-1}$
Mpc$^{-3}$.  Overall, the range of values found is in good agreement
with those derived in the current paper, and there is no clear reason
to favour either the high values we obtain with the \Ha GS luminosity
function, or the lower SDSS-derived value.

\section{Summary}

We have presented an analysis of the SF properties of
field galaxies within the local volume out to a recession velocity
limit of 3000~km~s$^{-1}$.  A parent sample of 863 galaxies with
well-defined Hubble type, apparent magnitude and diameter limits was
selected from the UGC, and used to calculate a $B$-band luminosity
function.  This was then populated with SF information from a
subsample of 327 galaxies, for which we have H$\alpha$ imaging,
firstly by calibrating a relationship between galaxy $B$-band
luminosity and SFR, and secondly by a Monte Carlo simulation of a
representative sample of galaxies, in which SF information was
randomly sampled from the observed subset.  The total SFR density of
the local Universe was found to be between 0.016 and 0.023~$M_{\odot}$
yr$^{-1}$ Mpc$^{-3}$ regardless of which of these two methods was
used, with the uncertainties being dominated by the internal
extinction correction used in converting measured H$\alpha$ fluxes to
SF rates.  If our internally derived $B$-band luminosity
function is replaced by one from the SDSS blue sequence, the SFR
densities are systematically reduced to about 60\% of the above
values.  This range of estimates is consistent with values from the
recent literature using a range of different SF indicators.

The Monte Carlo method was then used to calculate the contribution to
the total SFR density from galaxies of different luminosities and
Hubble $T$-types.  The largest contribution comes from bright galaxies
with $M_B\sim$--20~mag, and the total contribution from galaxies fainter
than $M_B=$--15.5~mag is less than 10\%. Almost 60\% of the SFR density
comes from galaxies of types Sb, Sbc or Sc; 9\% from galaxies earlier
than Sb and 33\% from galaxies later than Sc.  Finally, 75 - 80\% of the
total SF in the local Universe is shown to be occurring in
disk regions, defined as being $>$1~kpc from the centres of galaxies.

\begin{acknowledgements}
The Jacobus Kapteyn Telescope was operated on the island of La Palma
by the Isaac Newton Group in the Spanish Observatorio del Roque de los
Muchachos of the Instituto de Astrof\'isica de Canarias. This research
has made use of the NASA/IPAC Extragalactic Database (NED) which is
operated by the Jet Propulsion Laboratory, California Institute of
Technology, under contract with the National Aeronautics and Space
Administration. We thank the referee for a thoughtful and helpful
reading of the paper, which resulted in significant improvements in 
content and presentation.
\end{acknowledgements}

\bibliographystyle{bibtex/aa}
\bibliography{refs}
      

\begin{appendix}

\section{Morphological classification along the red and blue sequences}

With the advent of large imaging surveys, both local and at high
redshift, colour classification in particular dividing into a red and
blue sequence is often preferred to morphological classification.  The
is partly because of the reliability of determining galaxy colours
automatically compared to morphological characteristics for these
surveys, and partly because of the colour bimodality
\citep{strateva01}. Galaxies divide naturally into a red and blue
sequence when viewed in a colour-magnitude diagram (CMD) using $u-r$,
$g-r$ colours or similar \citep{blanton03broadband,baldry04}.  In
order to outline the connection between traditional morphological
classification using the E-S0-Sa-Sd-Im scheme and the colour
sequences, we used data from the Nearby Field Galaxy Survey (NFGS;
\citealt{jansen00}) and the SDSS Japanese Participation Group (JPG)
catalogue \citep{fukugita07}.

\begin{figure*}
\centerline{\includegraphics[width=0.49\textwidth]{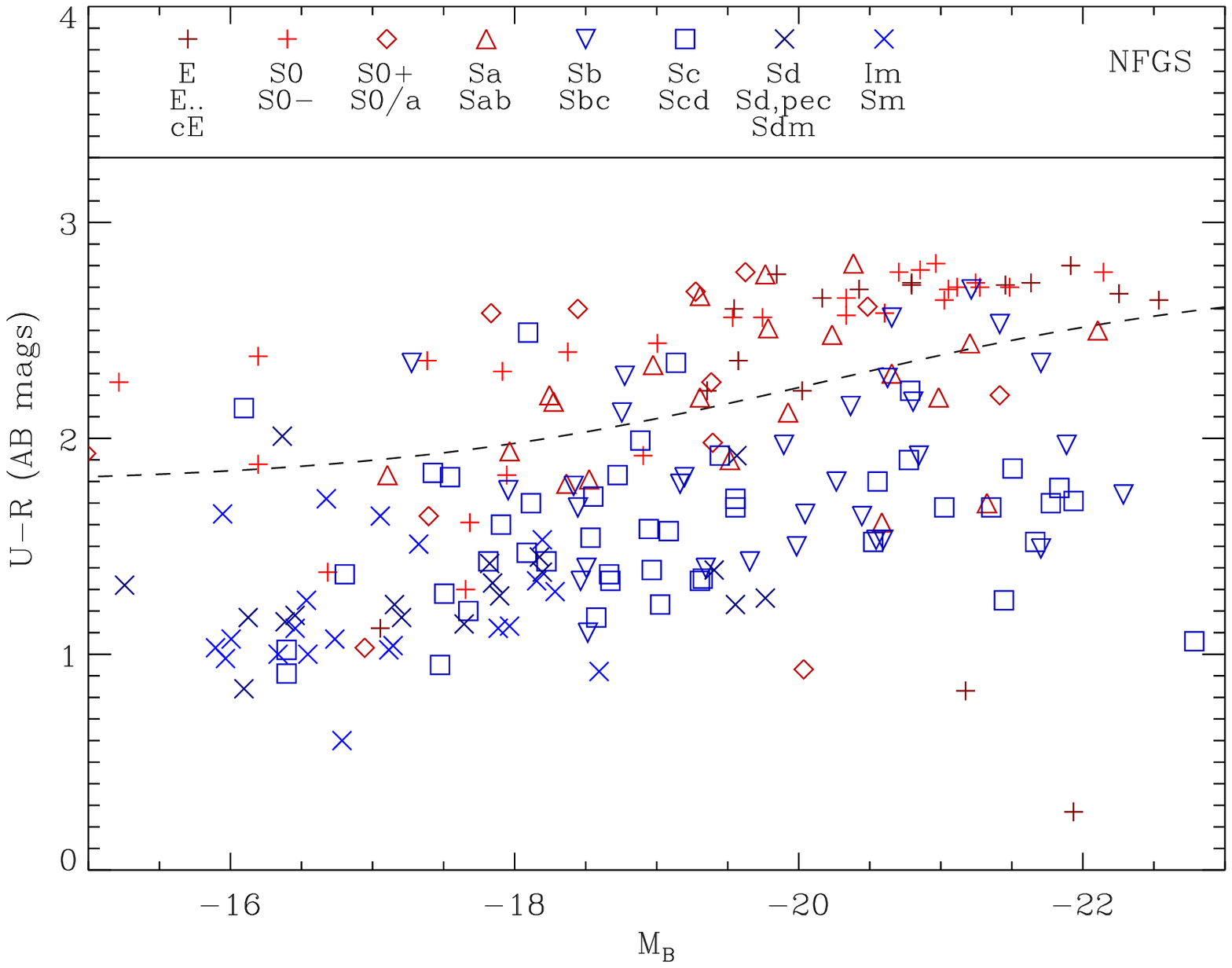}
            \includegraphics[width=0.49\textwidth]{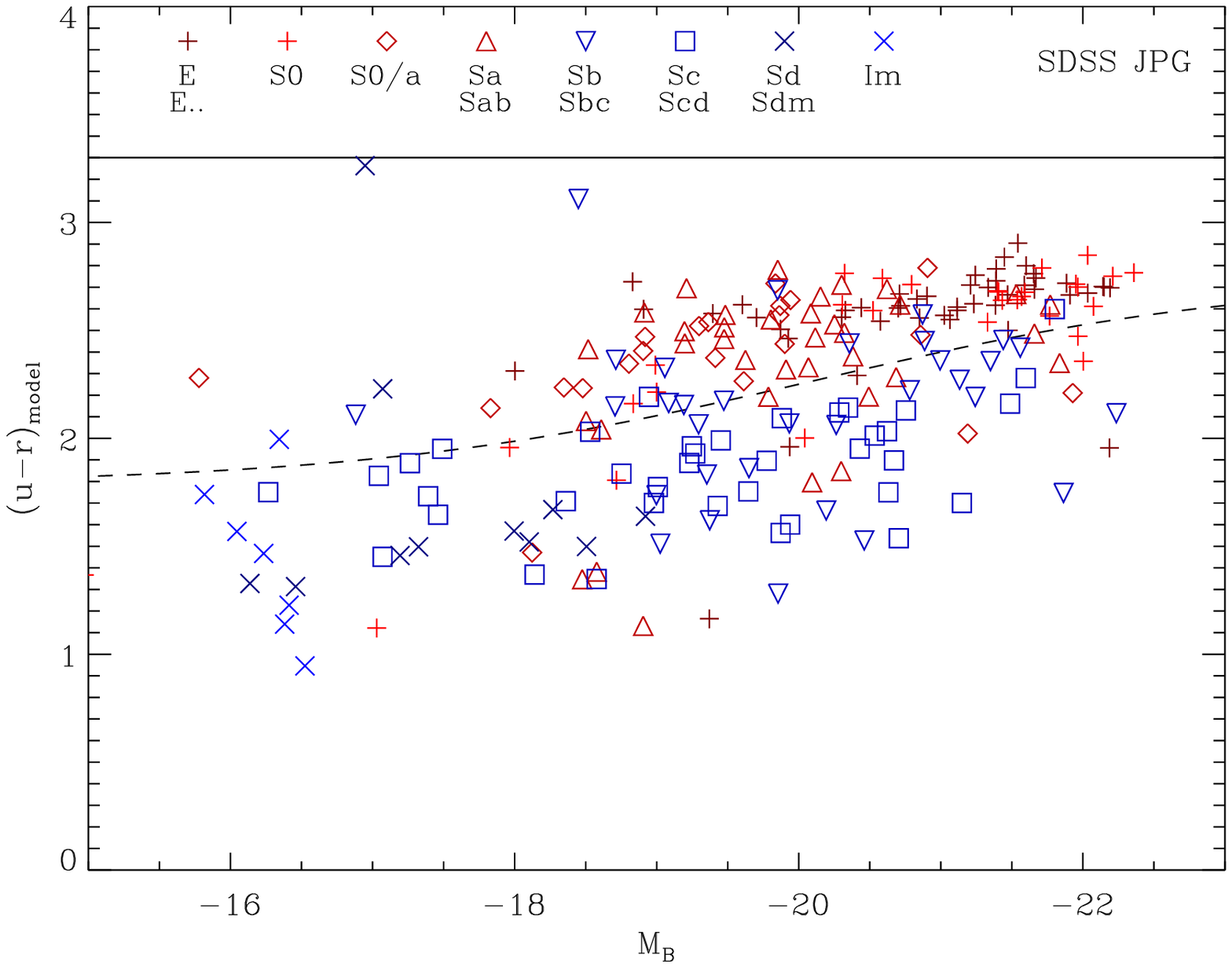}}
\caption{Morphological classifications in the CMD for galaxies. Left:
  NFGS sample. Right: A subset of the JPG sample.  The dashed line in
  both plots represents an optimal divider between the red and blue
  sequences determined from a larger SDSS sample.}
\label{fig:nfgs-jpg}
\end{figure*}

The NFGS includes $UBR$ surface photometry for 198 galaxies with a
median redshift of 0.01. The NFGS was selected from the
\citet{huchra83} catalogue, excluding the Virgo cluster, with sampling
to obtain a maximum of about 30 galaxies per 1-mag bin (blue absolute
magnitude).  We used the $U-R$ colours within the effective half-light
radius in $B$, converted to AB magnitudes ($+0.55$), and used the
absolute $B$-band magnitude converted to $H_0=70$~km~s$^{-1}$~Mpc$^{-1}$.
Figure~\ref{fig:nfgs-jpg} (left) shows the CMD for these data with
different symbols representing morphological classification in eight
groups.

The JPG catalogue includes SDSS photometry and eyeball classifications
for 2658 galaxies. This was reduced to 818 galaxies with $r<15.5$~mag and
$z<0.1$ (median redshift of 0.04). We used the $u-r$ `model'
rest-frame colour and a $B$-band absolute magnitude estimated using
the $g$ and $r$ magnitudes. SDSS model magnitudes are based on a
fitted profile in the $r$-band \citep{stoughton02}: effectively they
are centrally weighted colours.  Figure~\ref{fig:nfgs-jpg}(right)
shows the CMD for these data after sampling to obtain a maximum of 25
galaxies per 0.5-mag bin. The data are divided into eight
classification groups (note that an intermediate classification, e.g.\
Sab, represents an average between two or more classifiers.)

The dashed line in Figure~\ref{fig:nfgs-jpg} represents the best-fit
dividing line defined by \citet{baldry06}. This was determined from a
significantly larger SDSS sample. This appears to fall in about the
right place for the NFGS despite the different colour determination.
The faint-end of the red sequence is sparsely populated because these
are field samples. In order to compare late-type and blue-sequence
derived luminosity functions, we determined the fraction of galaxies
on the blue sequence ($f_b$) for the eight morphological groups.
This is shown in Table~\ref{tab:blue-fraction}. 

\begin{table}
\caption{Fraction of blue-sequence galaxies for different morphological types}
\label{tab:blue-fraction}
\begin{center}
\begin{tabular}{lrlrll} \hline
morphological & \multicolumn{2}{c}{NFGS}  & \multicolumn{3}{c}{SDSS JPG} \\
group         &  no.  &  $f_b$ &      no. &  $f_b$(A) & $f_b$(B)\\ \hline
E             &   17  &  0.24  &      201 &    0.04   &    0.03\\
S0            &	  28  &  0.21  &      114 &    0.18   &    0.11\\
S0/a          &	  12  &  0.42  &       73 &    0.19   &    0.22\\
Sa,ab         &   21  &  0.52  &      131 &    0.35   &    0.37\\
Sb,bc         &	  31  &  0.81  &      152 &    0.61   &    0.73\\
Sc,cd         &	  45  &  0.93  &      121 &    0.86   &    0.92\\
Sd,dm         &	  18  &  0.94  &       12 &    0.83   &    0.75\\
Im            &	  22  &  1.00  &       10 &    0.90   &    1.00\\ \hline
all           &  194  &  0.68  &      814 &    0.37   &    0.40\\ \hline
\end{tabular}
\end{center}			  		  
(A) using cut in $(u-r)_{\rm model}$ CMD, 
(B) using cut in $(g-r)_{\rm Petro}$ CMD.
\end{table}

When comparing $f_b$ it should first be noted that the NFGS and JPG
samples were selected using different criteria. For example, the JPG
sample is $r$-band magnitude limited and there was no explicit removal
of galaxies in high-density regions. Therefore, it is not surprising
that the overall $f_b$ is lower. Another factor is the higher median
redshift of the JPG which affects the quality of morphological
classification (at moderate imaging resolution of $\sim2''$).  For the
purposes of this paper the most important difference is the $f_b$
fraction for the Sb,bc galaxies because of their high contribution to
the SFR density, as found in section~5.2. This value is 0.61 for the
JPG sample (compared to 0.81 for the NFGS). This may partly account
for the lower SFR density derived from the blue-sequence SDSS
luminosity function (section~4).

\begin{figure*}
\centerline{\includegraphics[width=0.49\textwidth]{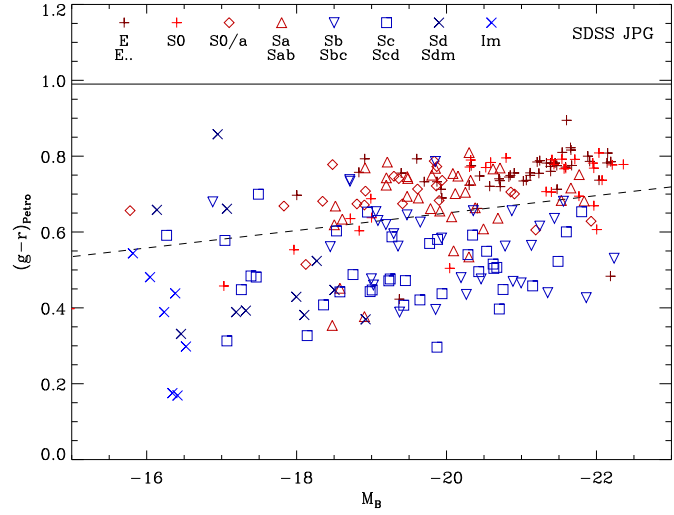}}
\caption{Morphological classifications in the CMD for galaxies. As per
  Figure~\ref{fig:nfgs-jpg} (right) except using $g-r$ Petrosian
  colours instead of $u-r$ model colours.  The dashed line represents
  an optimal divider determined by eye from a larger SDSS sample:
  $(g-r)_{\rm divide} = 0.65 - 0.023 (M_b + 20)$.}
\label{fig:gr-color-jpg}
\end{figure*}

One possible reason for a low $f_b$ value for the Sb-classified JPG
galaxies is the use of the $u-r$ model colour, which is centrally
weighted.  It is generally not reliable to use SDSS $u$-band
magnitudes derived from larger apertures because of Poisson noise and
sky subtraction, and an alternative is the $g-r$ Petrosian
colour. Figure~\ref{fig:gr-color-jpg} shows the $g-r$ CMD for the JPG
sample, and the $f_b$ values are also shown in the final column of
Table~\ref{tab:blue-fraction}. The $f_b$ value is now 0.73 for Sb,bc
galaxies indicating that aperture choice not surprisingly affects the
red and blue sequence classification of these galaxies.  Nevertheless,
a blue-sequence luminosity function can be regarded as being similar
to an Sb-or-later luminosity function because of the cross-over in
$f_b$ between Sa and Sb (considering the magnitude-limited JPG
sample).

\end{appendix}

\end{document}